# Eigenvector spatial filtering for large data sets: fixed and random effects approaches

Daisuke Murakami[1], Daniel A. Griffith[2]

[1]Department of Statistical Modeling, The Institute for Statistical Mathematics, 10-3 Midori-cho, Tachikawa, Tokyo, 190-8562, Japan

[2]School of Economic, Political and Policy Sciences, The University of Texas at Dallas, Richardson, Dallas, TX, 75083, USA

Correspondence: Daisuke Murakami, Department of Statistical Modeling, The Institute for Statistical Mathematics, 10-3 Midori-cho, Tachikawa, Tokyo, 190-8562, Japan
email: dmuraka@ism.ac.jp

*Eigenvector spatial filtering (ESF) is a spatial modeling approach, which has been applied in urban and regional studies, ecological studies, and so on. However, it is computationally demanding, and may not be suitable for large data modeling. The objective of this study is developing fast ESF and random effects ESF (RE-ESF), which are capable of handling very large samples. To achieve it, we accelerate eigen-decomposition and parameter estimation, which make ESF and RE-ESF slow. The former is accelerated by utilizing the Nyström extension, whereas the latter is by small matrix tricks. The resulting fast ESF and fast RE-ESF are compared with non-approximated ESF and RE-ESF in Monte Carlo simulation experiments. The result shows that, while ESF*

*and RE-ESF are slow for several thousand samples, fast ESF and RE-ESF require only several seconds for the samples. They also suggest that the proposed approaches effectively remove positive spatial dependence in the residuals with very small approximation errors when the number of eigenvectors considered is 200 or more. Note that these approaches cannot deal with negative spatial dependence. The proposed approaches are implemented in an R package "spmoran."*

**1. Introduction**

Large spatial data are rapidly increasing in accordance with the development of technologies relating to sensors (e.g., remote sensors, human sensors) and internet of things (IoT), which enables us to connect and accumulate a wide variety of spatial information (e.g., vehicle location, building energy use) through internetworking. In the era of BigData, fast statistical methods that are applicable for very large spatial samples are needed by both researchers and practitioners.

Statistical methods for spatial data have been developed mainly in geostatistics (e.g., Cressie, 1993), whose principal interest is spatial interpolation and other data-driven analyses (Anselin, 2010), and spatial econometrics (e.g., LeSage and Pace, 2009; Anselin and Rey, 2014), whose major interest is statistical inference in the presence of spatial dependence.

Unfortunately, estimation of geostatistic and spatial econometric models typically requires computational complexity of $O(n^3)$, which makes them intractable if the sample size *n* is large. A number of computationally efficient approximations have been developed in these study areas. They include likelihood approximations (e.g., Stein, Chi, and Welty, 2004; Griffith, 2004a; LeSage and Pace, 2007; Arbia, 2014), low rank approximations (e.g., Cressie and Johannesson, 2008; Hughes and Haran, 2013; Burden, Cressie, and Steel, 2015), spatial process approximations (e.g., Banerjee et al., 2008; Datta et al., 2016), and Gaussian Markov random field-based approximations (Lindgren, Rue, and Lindström, 2011) (See Sun, Li, and Genton, 2012 for review).

Among them, low rank approximation is a popular one. This approach attempts to describe spatial variation using a linear combination of *L* basis functions ($L << n$). Large *L* reduces approximation error, but increases computational complexity; hence, it is important to compress spatial variations efficiently, in a small number of basis functions.

In terms of compression efficiency, Moran's eigenvectors (or Moran bases: Griffith, 2003; Dray, Legendre, and Peres-Neto, 2006; Hughes and Haran, 2013) are useful. The first *L*-eigenvectors capture principal spatial dependent variations explained by the Moran coefficient (see Anselin and Rey, 1991). Moran's eigenvector-based spatial regression approach is called eigenvector spatial filtering (ESF: Griffith, 2003) in regional science, and also is called Moran's eigenvector maps (Borcard and Legendre, 2002; Dray, Legendre, and Peres-Neto, 2006) in ecology. Tiefelsdorf and Griffith (2007), Thayn and Simanis (2013), among others, demonstrate that ESF with a small number of eigenvectors (i.e., small *L*) greatly reduces model misspecification errors and increases model accuracy. Thus, ESF is a popular spatial model in applied studies (Pace, LeSage, and Zhu, 2013). Recently, Murakami and Griffith (2015) extend ESF, which is a fixed effects approach, to a random effects approach, which they call random effects ESF (RE-ESF). They show that RE-ESF estimates parameters with smaller estimation errors and less computation time than ESF.

Still, ESF and RE-ESF are slow, and not suitable for large samples (Dormann et al., 2007). Actually, (i) they require eigen-decomposition, whose computational complexity is $O(n^3)$. Furthermore, (ii) their parameter estimations also are computationally demanding; in particular, the classical ESF requires a stepwise eigenvector selection, which is very slow for large samples. Griffith (2000, 2015) proposes analytical solutions for the eigen-decomposition to solve problem (i). However, it is available only for regular square tessellation data, such as remote sensing data. Regarding (ii), Seya et al. (2015) show that stepwise selection can be substituted with the least absolute shrinkage and selection operator (LASSO) procedure (Tibshirani, 1996), which is a regularized estimation technique. Yet, LASSO also can be slow for a large dataset, as we demonstrate later.

The objective of this study is to accelerate ESF and RE-ESF for large samples. We perform a dimension reduction that approximates the first $L$ ($L < N$) eigen-pairs explaining positive spatial dependence (see Section 3.3). In other words, our approach is incapable of modeling negative spatial dependence (see Griffith, 2006).

The following sections are organized as follows. Section 2 introduces ESF and RE-ESF. Section 3 develops a fast approximation of Moran's eigenvectors to mitigate problem (i), and Section 4 proposes fast ESF and fast RE-ESF with fast parameter estimation to cope with problem (ii). Section 5 compares the proposed fast approximations with non-approximated spatial models. Section 6 compares fast approximations in a broader range of cases. Finally, Section 7 concludes our discussion.

## 2. Moran's eigenvector-based spatial approach

2.1. Moran's eigenvectors

The Moran coefficient (MC) quantifies spatial dependence in **y**, which is an $n \times 1$ vector of response variables, using the following equation:

$$MC[\mathbf{y}] = \frac{n}{\mathbf{1}'\mathbf{C}\mathbf{1}} \frac{\mathbf{y}'\mathbf{MCMy}}{\mathbf{y}'\mathbf{My}}, \tag{1}$$

where " ′ " denotes matrix transpose, $\mathbf{1}$ is a $n \times 1$ vector of ones. $\mathbf{M} = \mathbf{I} - \mathbf{11}'/n$ is a $n \times n$ centering matrix, where $\mathbf{I}$ is an identity matrix, and $\mathbf{C}$ is a $n \times n$ symmetric connectivity matrix whose diagonal entries are 0. Following Dray, Legendre, and Peres-Neto (2006), the $(i, j)$-th element of $\mathbf{C}$ is given by $c(s_i, s_j) = \exp(-d(s_i,s_j)/r)$, where $d(s_i,s_j)$ is the Euclidean distance between sample sites $s_i$ and $s_j$, and $r$ is the maximum length in the minimum spanning tree connecting the samples. Note that the exponential kernel can be replaced with a spherical kernel, Gaussian kernel, or any other positive-semi definite kernel (see Cressie, 1993).

Let us eigen-decompose the matrix $\mathbf{MCM}$ into $\mathbf{E}_{full}\mathbf{\Lambda}_{full}\mathbf{E}'_{full}$, where $\mathbf{E}_{full}$ is a $n \times n$ matrix whose $l$-th column, $\mathbf{e}_l$, equals the $l$-th eigenvector, and $\mathbf{\Lambda}_{full}$ is a $n \times n$ diagonal matrix with its $l$-th element being the $l$-th eigenvalue, $\lambda_l$. The MC of $\mathbf{e}_l$ is specified as follows:

$$MC[\mathbf{e}_l] = \frac{n}{\mathbf{1}'\mathbf{C}\mathbf{1}} \frac{\mathbf{e}'_l \mathbf{MCMe}_l}{\mathbf{e}'_l \mathbf{Me}_l} = \frac{n}{\mathbf{1}'\mathbf{C}\mathbf{1}} \frac{\mathbf{e}'_l \mathbf{E}_{full} \mathbf{\Lambda}_{full} \mathbf{E}'_{full} \mathbf{e}_l}{\mathbf{e}'_l \mathbf{e}_l},$$

$$= \frac{n}{\mathbf{1}'\mathbf{C}\mathbf{1}} \lambda_l. \tag{2}$$

Eq. (2) suggests that the eigenvectors are interpretable in terms of the MC. Specifically, the 1st eigenvector, $\mathbf{e}_1$, is the set of real numbers that has the largest MC value achievable by any set of real numbers for the spatial structure defined by $\mathbf{C}$; $\mathbf{e}_2$, is the set of real numbers that has the largest achievable MC value by any set that is orthogonal and uncorrelated with $\mathbf{e}_1$; and so forth, such that the $l$-th eigenvector, $\mathbf{e}_l$, is the set of real numbers that has the largest achievable MC value by any set that is orthogonal and uncorrelated with $\{\mathbf{e}_1, ..., \mathbf{e}_{l-1}\}$. Thus, $\mathbf{E}_{full} = \{\mathbf{e}_1, ..., \mathbf{e}_n\}$ provides all the possible distinct map pattern descriptions of latent spatial dependence, with each magnitude being indexed through its corresponding eigenvalue in $\{\lambda_1, ..., \lambda_n\}$ (Griffith, 2003).

2.2. Linear ESF models

The basic linear model of ESF is

$$\mathbf{y} = \mathbf{X}\boldsymbol{\beta} + \mathbf{E}\boldsymbol{\gamma} + \boldsymbol{\varepsilon}, \qquad \boldsymbol{\varepsilon} \sim N(\mathbf{0}, \sigma^2 \mathbf{I}), \qquad (3)$$

where $\mathbf{X}$ is a $n \times K$ matrix of the explanatory variables, $\mathbf{E}$ is a $n \times L$ matrix composed of a subset of $L$ eigenvectors ($L < n$) from $\mathbf{E}_{full}$, $\mathbf{0}$ is a $n \times 1$ vector of zeros, $\boldsymbol{\beta}$ and $\boldsymbol{\gamma}$ are vectors of coefficients, and $\sigma^2$ is a variance parameter.

Classical ESF considers $\boldsymbol{\gamma}$ as fixed, and defines $\mathbf{E}$ by a subset of $L$ eigenvectors chosen by a stepwise eigenvector selection, which is based on accuracy maximization or residual spatial dependence minimization (see Griffith and Chun, 2014, 2016). In contrast, RE-ESF assumes $\boldsymbol{\gamma}$ to be random such that

$$\boldsymbol{\gamma} \sim N(\mathbf{0}_L, \sigma_\gamma^2 \boldsymbol{\Lambda}(\alpha)). \qquad (4)$$

where $\mathbf{0}_L$ is a $L \times 1$ vector of zeros, and $\mathbf{\Lambda}(\alpha)$ is a $L \times L$ diagonal matrix whose $l$-th element is $\lambda_l(\alpha) = \left(\sum_l \lambda_l / \sum_l \lambda_l^\alpha\right)\lambda_l^\alpha$. $\alpha$ is an unknown multiplier determining the scale of the spatial dependence, and $\sigma_\gamma^2$ represents the variance. The RE-ESF model is identical with a Gaussian process after a rank reduction (Murakami and Griffith, 2015).

$L$ may be defined by the number of positive eigenvalues. In this case, all eigenvectors and eigenvalues describing positive spatial dependence are considered (Griffith, 2003). Recent literature on statistics confirms that this criterion successfully eliminates residual spatial dependence (e.g., Hughes and Haran, 2013; Johnson et al., 2015)

2.3. Properties of the ESF models

RE-ESF describes spatial dependence using $Var[\mathbf{E\gamma}] = \mathbf{E\gamma\gamma'E'} = \mathbf{E\Lambda}(\alpha)\mathbf{E'} = k\mathbf{C}^\alpha$, where $k$ is a constant. While the range parameter $r$ in $\mathbf{C}$ is assumed known,[1] the $\alpha$ parameter estimates the effective range, which is the distance that 95% of spatial dependence disappears (see Schechanberger and Gotway, 2004). Figure 1 illustrates the distance decay of spatial dependence being modeled by the matrix $\mathbf{C}^\alpha$ when $\alpha = 0.5, 1.0,$ and 2.0. This figure shows that the decay becomes fast when $\alpha$ is large, whereas it decays slowly when $\alpha$ is small. Thus, the parameter $\alpha$ allows estimation of the effective range even if the range parameter $r$ is fixed. Section 6 demonstrates the flexibility of this approach based on simulation experiments.

---

[1] Assumptions of a known spatial correlation matrix and known number of basis functions, $L$, are common in the literature about reduced rank spatial modeling (e.g., Cressie and Johannesson, 2008; Hughes and Haran, 2013; Burden et al., 2015).

[Figure 1 around here]

Besides, ESF and RE-ESF, which use principal eigenvectors, are robust to the choice of the **C** matrix because of the following reasons: the principal eigenvectors tend to be quite similar even if **C** is changed (see Griffith and Peres-Neto, 2006); the eigenvectors are independent of the scale of the spatial dependence, or the effective range, as illustrated in Figure 1.

## 3. An approximation of the Moran's eigenfunctions

This section approximates **E** and **Λ**, using the property that the eigenvectors and the eigenvalues of $\mathbf{MC^+M} = \mathbf{M(C+I)M}$ are given by **E** and $\mathbf{\Lambda}+\mathbf{I}_L$, respectively, where $\mathbf{I}_L$ is a $L \times L$ identity matrix (see Griffith, 2003)[2]. Specifically, after imposing assumptions

---

[2] Use of this property is needed because the Nyström extension, which we use in Section 3.3, is only for positive semi-definite matrix, while **MCM** is by definition an indefinite

in Section 3.1, Section 3.2 analyzes properties of **MC⁺M**, and Section 3.3 approximates

**E** and **Λ** based on the result.

*3.1. Assumptions*

We approximate the Moran's eigenfunctions for *n* samples using eigenfunctions defined for *L* knots, which are distributed across a target area (see Figure 3 for an illustration). Hereafter, spatial coordinates for the *n* samples are denoted by $s_i | i \in \{1,...n\}$, whereas those for the *L* knots are represented by $s_I | I \in \{1,... L\}$ (i.e., indices for the knots are given by uppercase letters).

The following assumptions are imposed:

$$\frac{1}{n}\sum_{j=1}^{n} c(s_i, s_j) \approx \frac{1}{L}\sum_{I=1}^{L} c(s_I, s_j), \tag{5}$$

$$\frac{1}{n}\sum_{j=1}^{n} c(s_i, s_j) \approx \frac{1}{L}\sum_{J=1}^{L} c(s_i, s_J), \tag{6}$$

$$\frac{1}{n^2}\sum_{i=1}^{n}\sum_{j=1}^{n} c(s_i, s_j) \approx \frac{1}{L^2}\sum_{I=1}^{L}\sum_{J=1}^{L} c(s_I, s_J), \tag{7}$$

where $s_I$ denotes the *I*-th knot, and $c(s_i, s_j) = \exp(-d(s_i, s_j)/r)$ (see Section 2.1). Eqs. (5),

---

matrix. **MC⁺M** is necessarily positive semi-definite as long as the (*i*, *j*)-th element of **C** is given by exp(-$d(s_i, s_j)/r$). The exponential kernel is replaced with other positive semi-definite kernels if only the elements of **C** are given using those kernels in which the range parameter is estimated a priori, as explained in Section 2.1. Namely, other kernels are usable without any change of our methodology. Appendix B demonstrates through a simulation experiment that our approach works well even when other kernels are used.

(6), and (7) assume that the average connectivity among sample sites is approximated by the average connectivity among knots. This assumption holds if the $L$ knots have similar distributional properties with the $n$ samples.

*3.2. A property of the matrix $\mathbf{MC^+M}$*

This section aims to associate the $n \times n$ matrix $\mathbf{MC^+M}$ with the $L \times L$ connectivity matrix among $L$ knots, $\mathbf{M}_L \mathbf{C}^+_L \mathbf{M}_L$, where $\{\mathbf{C}^+_L (L \times L)$, and $\mathbf{M}_L (L \times L)\}$ are defined similar to $\{\mathbf{C}^+, \mathbf{M}\}$.

To achieve this outcome, we first write the $(i, j)$-th element of $\mathbf{MC^+M} = \mathbf{C}^+ - \mathbf{11'C^+}/n - \mathbf{C^+11'}/n + \mathbf{11'C^+11'}/n^2$ as

$$c_{MCM}(s_i, s_j) = c(s_i, s_j) - \frac{1}{n}\sum_{i=1}^{n} c(s_i, s_j) - \frac{1}{n}\sum_{j=1}^{n} c(s_i, s_j) + \frac{1}{n^2}\sum_{i=1}^{n}\sum_{j=1}^{n} c(s_i, s_j). \qquad (8)$$

Eq. (8) is approximated by substituting Eqs. (5), (6), and (7) into Eq. (8), as follows:

$$c_{MCM}(s_i, s_j) \approx c(s_i, s_j) - \frac{1}{L}\sum_{J=1}^{L} c(s_i, s_J) - \frac{1}{L}\sum_{I=1}^{L} c(s_I, s_j) - \frac{1}{L^2}\sum_{I=1}^{L}\sum_{J=1}^{L} c(s_I, s_J). \qquad (9)$$

Based on Eq. (9), the spatial connectivity between knots $s_I$ and $s_J$ is

$$c_{MCM}(s_I, s_J) \approx c(s_I, s_J) - \frac{1}{L}\sum_{J=1}^{L} c(s_I, s_J) - \frac{1}{L}\sum_{I=1}^{L} c(s_I, s_J) - \frac{1}{L^2}\sum_{I=1}^{L}\sum_{J=1}^{L} c(s_I, s_J), \qquad (10)$$

while the connectivity between a sample site $s_i$ and a knot $s_J$ is

$$c_{MCM}(s_i, s_J) \approx c(s_i, s_J) - \frac{1}{L}\sum_{J=1}^{L} c(s_i, s_J) - \frac{1}{L}\sum_{I=1}^{L} c(s_I, s_J) - \frac{1}{L^2}\sum_{I=1}^{L}\sum_{J=1}^{L} c(s_I, s_J). \quad (11)$$

The $L \times L$ matrix whose $(I, J)$-th element is given by Eq. (10) results in $\mathbf{M}_L \mathbf{C}_L^+ \mathbf{M}_L$. The $1 \times L$ vector whose $J$-th element is given by Eq. (11) yields

$$\mathbf{c}_{MCM}(s_i) = \mathbf{c}_L(s_i) - \mathbf{c}_L(s_i)\mathbf{1}_L\mathbf{1}_L' / L - \mathbf{1}_L'\mathbf{C}_L^+ / L + \mathbf{1}_L'\mathbf{C}_L^+\mathbf{1}_L\mathbf{1}_L' / L^2$$

$$= \mathbf{c}_L(s_i)(\mathbf{I}_L - \mathbf{1}_L\mathbf{1}_L' / L) - \mathbf{1}_L'\mathbf{C}_L^+ / L(\mathbf{I}_L - \mathbf{1}_L\mathbf{1}_L' / L)$$

$$= (\mathbf{c}_L(s_i) - \mathbf{1}_L'\mathbf{C}_L^+ / L)(\mathbf{I}_L - \mathbf{1}_L\mathbf{1}_L' / L), \quad (12)$$

where $\mathbf{c}_L(s_i)$ is a $1 \times L$ vector whose $J$-th element is $c(s_i, s_J) = \exp(-d(s_i,s_J)/r)$, and $\mathbf{1}_L$ is a $L \times 1$ vector of ones. Based on the preceding discussion, under assumptions Eqs. (5) - (7), spatial connectivity among the $L + n$ (knots + sample) sites is described as follows:

$$\begin{pmatrix} \mathbf{M}_L \mathbf{C}_L^+ \mathbf{M}_L & \mathbf{M}_L[(\mathbf{C}_{nL} - \mathbf{1} \otimes (\mathbf{1}_L'\mathbf{C}_L^+ / L)]' \\ [\mathbf{C}_{nL} - \mathbf{1} \otimes (\mathbf{1}_L'\mathbf{C}_L^+ / L)]\mathbf{M}_L & \mathbf{M}\mathbf{C}^+\mathbf{M} \end{pmatrix}, \quad (13)$$

with its $(i, j)$-th element being given by Eq. (8). $\otimes$ denotes the Kronecker product operator, and $\mathbf{C}_{nL}$ is a $n \times L$ matrix whose $(i, J)$-th element is $c(s_i, s_J)$.

Section 3.3 reveals how to approximate $\mathbf{E}$ and $\mathbf{\Lambda}$ using Eq. (13).

*3.3. An eigenfunction approximation using the Nyström extension*

Let $\begin{pmatrix} \mathbf{G} & \mathbf{H}_{12}' \\ \mathbf{H}_{12} & \mathbf{H}_{22} \end{pmatrix}$ be a $(G + H) \times (G + H)$ matrix for which $\mathbf{G}$, $\mathbf{H}_{12}$, and $\mathbf{H}_{22}$ are matrixes whose sizes are $G \times G$, $H \times G$, and $H \times H$ ($G \ll H$). Given $\mathbf{G}$ and $\mathbf{H}_{12}$, the least squares solution for $\mathbf{H}_{22}$ is given by $\hat{\mathbf{H}}_{22} = \mathbf{H}_{12}\mathbf{G}^{-P}\mathbf{H}'_{12}$, where $\mathbf{G}^{-P}$ is the pseudo-inverse of $\mathbf{G}$. Using this fact, the Nyström extension (Drineas and Mahoney 2005) approximates the first $G$ eigenvectors of $\mathbf{H}_{22}$ with $\hat{\mathbf{E}}_{H_{22}} = \mathbf{H}_{12}\mathbf{E}_G\mathbf{\Lambda}_G^{-1}$, where $\mathbf{G} = \mathbf{E}_G\mathbf{\Lambda}_G\mathbf{E}_G'$. The corresponding eigenvalues are given by the diagonals of $((G+H)/G)\mathbf{\Lambda}_G$.

In our case, $\mathbf{E}$ and $\mathbf{\Lambda} + \mathbf{I}_L$, which are the first eigenfunctions for $\mathbf{MC^+M}$, are approximated using Eq. (13), as follows:

$$\hat{\mathbf{E}} = [\mathbf{C}_{nL} - \mathbf{1} \otimes (\mathbf{1}'_L \mathbf{C}_L^+ / L)]\mathbf{E}_L(\mathbf{\Lambda}_L + \mathbf{I}_L)^{-1}, \qquad (14)$$

$$\hat{\mathbf{\Lambda}} + \mathbf{I}_L = \frac{L+n}{L}(\mathbf{\Lambda}_L + \mathbf{I}_L), \qquad (15)$$

where $\mathbf{M}_L\mathbf{C}^+_L\mathbf{M}_L = \mathbf{E}_L(\mathbf{\Lambda}_L + \mathbf{I}_L)\mathbf{E}_L'$. $\mathbf{M}_L\mathbf{E}_L = \mathbf{E}_L$, which holds because of the zero means of the eigenvectors in $\mathbf{E}_L$, is used to obtain Eq. (14). Suppose that $\hat{\mathbf{E}} = [\hat{\mathbf{e}}_1, \ldots \hat{\mathbf{e}}_l, \ldots \hat{\mathbf{e}}_L]$; then Eq. (14) implies

$$\hat{\mathbf{e}}_l = \frac{1}{\lambda_{(L),l}+1}[\mathbf{C}_{nL} - \mathbf{1} \otimes (\mathbf{1}'_L \mathbf{C}_L^+ / L)]\mathbf{e}_{(L),l}, \qquad (16)$$

where $\mathbf{e}_{(L),l}$ and $\lambda_{(L),l}$ are the $l$-th eigenvector and eigenvalue of matrix $\mathbf{C}_L$, respectively.

Recall that $\mathbf{E}$ and $\mathbf{\Lambda}$ are eigenfunctions of $\mathbf{MCM}$, which we want to approximate. $\mathbf{E}$ is already provided by Eq. (14). $\mathbf{\Lambda}$ also is obtained based on Eq. (15), as follows:

$$\hat{\mathbf{\Lambda}} = \frac{L+n}{L}(\mathbf{\Lambda}_L + \mathbf{I}_L) - \mathbf{I}_L, \qquad (17)$$

Eqs. (14) and (17) estimate $\mathbf{E}$ and $\mathbf{\Lambda}$ without explicitly handling $\mathbf{MCM}$. In other words, they decrease the computational complexity from $O(n^3)$ to $O(L^3)$, which is required to eigen-decompose $\mathbf{M}_L\mathbf{C}^+_L\mathbf{M}_L$. Its cost reduction is substantial as long as $L << n$.

The remaining problem is how to distribute the $L$ knots. Following Zhang and Kwok (2010), we use $k$-means clustering centers as the knots for the Nyström extension. As they show, the computational complexity of $k$-mean clustering is only $O(n)$. In addition, use of the $k$-means centers effectively reduces approximation errors.

Figure 2 plots 200 knots extracted from 9,215 official sample sites of 2010 residential land prices in the Tokyo metropolitan area, Japan. The 200 knots should effectively cover the sample space. Figure 3 plots the 1st, 10th, and 100th eigenvectors using the knots. As with the usual Moran's eigenvectors, the 1st eigenvector has a global map pattern, the 10th has a moderate-scale map pattern, and the 100th has a local map pattern. We verified that our approximation successfully captures spatial characteristics of Moran's eigenvectors.

**[Figure 2 around here]**

**[Figure 3 around here]**

## 4. Fast ESF and RE-ESF

Section 4.1 introduces models for the fast ESF and RE-ESF. Subsequently,

Sections 4.2 and 4.3 respectively clarify how to estimate the fast ESF and RE-ESF models.

*4.1. The model*

The basic linear model for fast ESF is established by replacing **E** in Eq. (3) with $\hat{\mathbf{E}}$, which is derived in Section 3, as follows:

$$\mathbf{y} = \mathbf{X}\boldsymbol{\beta} + \hat{\mathbf{E}}\boldsymbol{\gamma} + \boldsymbol{\varepsilon}, \qquad \boldsymbol{\varepsilon} \sim N(\mathbf{0}, \sigma^2 \mathbf{I}), \qquad (18)$$

where the fast ESF considers **γ** as fixed. In contrast, the fast RE-ESF considers **γ** as a random coefficients vector obeying the estimated version of expression (4),

$$\boldsymbol{\gamma} \sim N(\mathbf{0}_L, \sigma_\gamma^2 \hat{\boldsymbol{\Lambda}}(\alpha)), \qquad (19)$$

where $\hat{\boldsymbol{\Lambda}}$ is the estimated matrix.

This study focuses on positive spatial dependence, which is dominant in many cases in regional science. Hence, eigenvectors, $\hat{\mathbf{e}}_l$, explaining positive spatial dependence (i.e., $\hat{\lambda}_l > 0$; see Eq. (2)) are considered in the subsequent analysis.

*4.2. Estimation: fast ESF*

Although the standard ESF employs a stepwise eigenvector selection procedure, this selection procedure is disappointedly slow if *n* is large (Dormann et al., 2007). This study proposes the following simple alternatives to estimate fast ESF: (i) without selection (Pace, LeSage, and Zhu, 2013); and, (ii) correlation-based screening (see Fan and Lv 2008). This first alternative simply selects all eigenvectors satisfying $\hat{\lambda}_l > 0$. This approach is acceptable when $n$ is large and the loss of degrees of freedom is small even without the eigenvector selection. The second alternative selects eigenvectors whose correlation coefficients with **y** exceed a preset threshold in absolute value.

*4.3. Estimation: fast RE-ESF*

This study proposes the following estimation procedure for fast RE-ESF:

1: $\mathbf{M}_{XX} = \mathbf{X}'\mathbf{X}$, $\mathbf{M}_{EX} = \hat{\mathbf{E}}'\mathbf{X}$, $\mathbf{M}_{EE} = \hat{\mathbf{E}}'\hat{\mathbf{E}}$, $\mathbf{m}_{Xy} = \mathbf{X}'\mathbf{y}$, $\mathbf{m}_{Ey} = \hat{\mathbf{E}}'\mathbf{y}$, and $m_{yy} = \mathbf{y}'\mathbf{y}$ are calculated.

2: $\boldsymbol{\theta} \in \{\alpha, \sigma_\gamma^2\}$ are numerically estimated by maximizing their profile log-likelihood (see Murakami and Griffith 2015), Eq. (20), in which $\mathbf{X}'\mathbf{X}$, $\hat{\mathbf{E}}'\mathbf{X}$, $\hat{\mathbf{E}}'\hat{\mathbf{E}}$, $\mathbf{X}'\mathbf{y}$, $\hat{\mathbf{E}}'\mathbf{y}$, and $\mathbf{y}'\mathbf{y}$ are replaced following the first step:

$$loglik_R(\boldsymbol{\theta}) = -\frac{1}{2}\log\left\|\begin{matrix}\mathbf{M}_{XX} & \mathbf{M}'_{EX}\hat{\mathbf{V}}(\boldsymbol{\theta}) \\ \hat{\mathbf{V}}(\boldsymbol{\theta})\mathbf{M}_{EX} & \hat{\mathbf{V}}(\boldsymbol{\theta})\mathbf{M}_{EE}\hat{\mathbf{V}}(\boldsymbol{\theta}) + \mathbf{I}_L\end{matrix}\right\| - \frac{n-K}{2}\left[1+\log\left(\frac{2\pi(\hat{\boldsymbol{\varepsilon}}'\hat{\boldsymbol{\varepsilon}} + \hat{\mathbf{u}}'\hat{\mathbf{u}})}{n-K}\right)\right], \qquad (20)$$

where $\boldsymbol{\gamma} = \hat{\mathbf{V}}(\boldsymbol{\theta})\mathbf{u}$, $\hat{\mathbf{V}}(\boldsymbol{\theta}) = \sigma_\gamma \hat{\boldsymbol{\Lambda}}(\alpha)^{1/2}$, and $\mathbf{u} \sim N(\mathbf{0}_L, \mathbf{I}_L)$, and

$$\hat{\boldsymbol{\varepsilon}}'\hat{\boldsymbol{\varepsilon}} = m_{yy} - 2[\boldsymbol{\beta}', \mathbf{u}'] \begin{bmatrix} \mathbf{m}_{Xy} \\ \hat{\mathbf{V}}(\boldsymbol{\theta})\mathbf{m}_{Ey} \end{bmatrix} + [\boldsymbol{\beta}', \mathbf{u}'] \begin{bmatrix} \mathbf{M}_{XX} & \mathbf{M}'_{EX}\hat{\mathbf{V}}(\boldsymbol{\theta}) \\ \hat{\mathbf{V}}(\boldsymbol{\theta})\mathbf{M}_{EX} & \hat{\mathbf{V}}(\boldsymbol{\theta})\mathbf{M}_{EE}\hat{\mathbf{V}}(\boldsymbol{\theta}) \end{bmatrix} \begin{bmatrix} \boldsymbol{\beta} \\ \mathbf{u} \end{bmatrix}, \quad (21)$$

$$\begin{bmatrix} \hat{\boldsymbol{\beta}} \\ \hat{\mathbf{u}} \end{bmatrix} = \begin{bmatrix} \mathbf{M}_{XX} & \mathbf{M}'_{EX}\hat{\mathbf{V}}(\boldsymbol{\theta}) \\ \hat{\mathbf{V}}(\boldsymbol{\theta})\mathbf{M}_{EX} & \hat{\mathbf{V}}(\boldsymbol{\theta})\mathbf{M}_{EE}\hat{\mathbf{V}}(\boldsymbol{\theta}) + \mathbf{I}_L \end{bmatrix}^{-1} \begin{bmatrix} \mathbf{m}_{Xy} \\ \hat{\mathbf{V}}(\boldsymbol{\theta})\mathbf{m}_{Ey} \end{bmatrix}. \quad (22)$$

3: $\boldsymbol{\beta}$ and $\sigma^2$ are estimated by substituting estimated $\boldsymbol{\theta} \in \{\alpha, \sigma_\gamma^2\}$ into Eqs. (22) and (23), respectively[3]:

$$\hat{\sigma}^2 = \frac{1}{n-K} \hat{\boldsymbol{\varepsilon}}'\hat{\boldsymbol{\varepsilon}}, \quad (23)$$

4: The covariance matrix of the coefficient estimates is evaluated as

$$Var\begin{bmatrix} \hat{\boldsymbol{\beta}} \\ \hat{\mathbf{u}} \end{bmatrix} = \hat{\sigma}^2 \begin{bmatrix} \mathbf{M}_{XX} & \mathbf{M}'_{EX}\hat{\mathbf{V}}(\boldsymbol{\theta}) \\ \hat{\mathbf{V}}(\boldsymbol{\theta})\mathbf{M}_{EX} & \hat{\mathbf{V}}(\boldsymbol{\theta})\mathbf{M}_{EE}\hat{\mathbf{V}}(\boldsymbol{\theta}) + \mathbf{I}_L \end{bmatrix}^{-1}, \quad (24)$$

in which the diagonal entries are available to test the significance of $\hat{\boldsymbol{\beta}}$ and $\hat{\mathbf{u}}$.

Interestingly, owning to the first step, any matrices and vectors whose size depends on *n* do not appear after the second step. As a result, the computational complexity to evaluate the profile log-likelihood is only $O((K+L)^3)$. The computational time for the optimization of $\boldsymbol{\theta}$ is independent of the sample size.

In summary, this section proposes the fast ESF and RE-ESF approaches, which are applicable for large samples. Approximated eigenvectors and eigenvalues, which are used in these approaches, are calculated as derived in Section 3, and their parameters are estimated as explained in Section 4.

---

[3] $\boldsymbol{\gamma}$ is estimated by $\hat{\boldsymbol{\gamma}} = \hat{\mathbf{V}}(\boldsymbol{\theta})\hat{\mathbf{u}}$.

## 5. A simulation study

This section summarizes results from a Monte Carlo simulation experiment comparing the proposed fast approaches with standard ESF/RE-ESF approaches, in terms of computational time and parameter estimation accuracy. Section 5.1 outlines the setting of the experiment. Sections 5.2 compares the linear regression model (LM), ESF, and RE-ESF specifications.

5.1. An outline of the study

Simulated data are generated from Eq. (25):

$$\mathbf{y} = \beta_0 \mathbf{1} + \mathbf{x}_1 \beta_1 + \mathbf{x}_2 \beta_2 + \mathbf{E}\boldsymbol{\gamma} + \boldsymbol{\varepsilon}, \qquad (25)$$

$$\boldsymbol{\gamma} \sim N(\mathbf{0}, \sigma_\gamma^2 \boldsymbol{\Lambda}(1)), \qquad \boldsymbol{\varepsilon} \sim N(\mathbf{0}, \mathbf{I}),$$

where $\mathbf{x}_1$ and $\mathbf{x}_2$ are vectors of explanatory variables, and $\beta_0$, $\beta_1$, and $\beta_2$ are coefficients. Spatial coordinates of the simulated data are generated from the standard normal distribution[4]. $\mathbf{E}$ and $\boldsymbol{\Lambda}$ are given by non-approximated eigenvectors and eigenvalues corresponding to $\lambda_l > 0$, respectively.

---

[4] Use of the standard normal distribution implicitly assumes fewer samples in the suburbs of a target area, a likely feature of most regional data.

It is known that spatial dependence variation in explanatory variables can confound with residual spatial dependence, and make parameter estimates unstable (spatial confounder: e.g., Paciorek, 2010; Hodges and Reich, 2010; Hughes and Haran, 2013). With this in mind, $\mathbf{x}_k \in \{\mathbf{x}_1, \mathbf{x}_2\}$ is generated with Eq. (26):

$$\mathbf{x}_k = \mathbf{E}\boldsymbol{\gamma}_{x(k)} + \boldsymbol{\varepsilon}_{x(k)}, \qquad (26)$$

$$\boldsymbol{\gamma} \sim N(\mathbf{0}, \sigma^2_{\gamma(x(k))}\boldsymbol{\Lambda}(1)) \qquad \boldsymbol{\varepsilon}_{x(k)} \sim N(\mathbf{0}, (1-\sigma_{\gamma(x(k))})^2\mathbf{I})$$

where $\sigma_{\gamma(x(k))}$ is the rate of spatial dependent variation accounting for the total variation in $\mathbf{x}_1$ or $\mathbf{x}_2$.

Table 1 summarizes models we compare. These include the linear regression (LM), standard ESF-adjusted $R^2$ maximization-based forward eigenvector selection (E$_{step}$), RE-ESF (RE), and ESF-LASSO (E$_{lasso}$; Seya et al. 2015) specifications, and our proposed approximations. An ESF-LASSO first selects eigenvectors with the LASSO, and then applies OLS to the ESF model with the selected eigenvectors. Seya et al. (2015) report that the ESF-LASSO is a fast alternative to standard ESF.

The other models entail fast approximations. fE$_{100}$ and fE$_{200}$ use the first 100 and 200 approximated eigenvectors (i.e., $L = 100$ and 200), respectively, whose corresponding eigenvalues are positive. fE$_{100*}$ and fE$_{200*}$ further exclude eigenvectors whose correlation coefficients with $\mathbf{y}$ are below 0.01. fRE$_{50}$, fRE$_{100}$, and fRE$_{200}$ are fast RE-ESF models with 50, 100, and 200 approximated eigenvectors, respectively.

**[Table 1 around here]**

We assume a sample size of 5,000. True regression coefficients are set as follows: $\beta_0 = 1.0$, $\beta_1 = 2.0$, and $\beta_2 = -0.5$. Their estimates and standard errors are evaluated by varying $\sigma_\gamma \in \{0.5, 1.0, 2.0\}$ and $\sigma_{\gamma(x(k))} \in \{0, 0.6\}$. Each of the six cases have 200 replications. All of our calculations are implemented in a Windows 10 64-bit system with 48 GB of memory, and coded using R (version 3.3.0).

5.2. Result

Because results for $\beta_1$ and $\beta_2$ are very similar, we report results for $\beta_1$ only. The accuracy of estimates is evaluated by the bias and the root mean squared error (RMSE), which are formulated as follows:

$$Bias(\hat{\beta}_1) = \frac{1}{200}\sum_{iter=1}^{200}(\hat{\beta}_1^{iter} - \beta_1), \qquad (27)$$

$$RMSE(\hat{\beta}_1) = \sqrt{\frac{1}{200}\sum_{iter=1}^{200}(\hat{\beta}_1^{iter} - \beta_1)^2}, \qquad (28)$$

where *iter* denotes iteration number, $\hat{\beta}_1^{iter}$ is the estimate of $\beta_1$ given in the *iter*-th iteration.

Accuracy of the standard error estimate, $se[\hat{\beta}_1]$, also is important to appropriately test the statistical significance of $\beta_1$. To compare the accuracy of $se[\hat{\beta}_1]$, whose true value changes across iterations, the root mean squared percentage error (RMSPE) is evaluated. RMSPE is formulated as

$$RMSPE\,(se[\hat{\beta}_1]) = \sqrt{\frac{1}{200}\sum_{iter=1}^{200}\left(\frac{se[\hat{\beta}_1^{iter}] - se[\beta_1^{iter}]}{se[\beta_1^{iter}]}\right)^2}, \qquad (29)$$

where $se[\hat{\beta}_1^{iter}]$ is $se[\hat{\beta}_1]$ obtained in the *iter*-th iteration, and $se[\beta_1^{iter}]$ is the standard error, which is estimated by substituting the true parameter values into the true model, Eq. (25). In other words, Eq. (29) evaluates how accurately the reduced rank models approximate the standard error of the true model.

Tables 2 summarizes the bias in $\hat{\beta}_1$. The biases in fE$_{100*}$ and fE$_{200*}$ are larger than those for the other models. fE$_{200*}$ has the largest bias of -0.056 when $\sigma_{\gamma(x(k))}$=0.6 and $\sigma_\gamma$=2.0, which assumes strong spatial dependence in both explanatory variables and residuals. By contrast, biases in other spatial models are less than 0.011 in absolute value, which are quite small relative to the true value of $\beta_1 = 1.0$.

**[Table 2 around here]**

Tables 3 and 4 summarize the RMSEs for $\beta_1$ and the RMSPE for $se[\beta_1]$,

respectively. As expected, the LM has a large $RMSE(\hat{\beta}_1)$ and $RMSPE(se[\hat{\beta}_1])$. This outcome confirms that ignoring spatial dependence results in an erroneous conclusion. By contrast, these tables show that the ESF and RE-ESF specifications furnish $\hat{\beta}_1$ with a small RMSE, and $se[\hat{\beta}_1]$ with a small RMSPE, regardless of whether or not approximated eigenvectors are used.

[Table 3 around here]

[Table 4 around here]

Among ESF specifications, $RMSE(\hat{\beta}_1)$ and $RMSPE(se[\hat{\beta}_1])$ in fE$_{200}$ are comparable with the standard ESF, E$_{step}$ (or E$_{lasso}$). Actually, the RMSEs of E$_{step}$ and fE$_{200}$ are less than 0.027 across all cases, whereas the RMSEs of LM have a maximum of 0.368. fE$_{200}$ also is comparable with E$_{step}$ in terms of $RMSPE(se[\hat{\beta}_1])$. Thus, approximation errors in fE$_{200}$ are found to be small.

Among RE-ESF specifications, $RMSE(\hat{\beta}_1)$ and $RMSPE(se[\hat{\beta}_1])$ decrease as the number of eigenvectors, $L$, increases. When $L = 200$ (i.e., fRE$_{200}$), they are essentially the same as for the original RE-ESF model. The first 200 eigenvectors might be sufficient in a fast RE-ESF specification. The RMSPEs in fRE$_{200}$ are less than 0.022 across all cases, while those for fE$_{200}$ are less than 0.129. The RE-ESF specifications tend to outperform the ESF specifications.

Figure 4 portrays a comparison of computation times. As expected, E$_{step}$ was the slowest due to the eigenvector selection involved. Although E$_{lasso}$ and RE do not require such a selection, they are still slow because of the eigen-decomposition involved. In contrast, fE$_{200}$ and fRE$_{200}$ take only several seconds for $n = 10,000$, and only several minutes for $n = 500,000$.

[Figure 4 around here]

## 6. An additional simulation study: fast ESF and RE-ESF in cases with larger $n$

The previous section finds that fE$_{200}$ and fRE$_{200}$ accurately estimate coefficients computationally efficiently. Yet, it is unclear to what extent the result is valid for larger sample sizes. This validity might be lost when $n$ is large and the range parameter, $r$, which is given a priori (see Section 2.1), is misspecified. This section examines these situations with an additional simulation experiment. Here, the standard ESF, E$_{step}$, is not used because it becomes computationally intractable when $n \geq 10,000$.

6.1. An outline of the study

This section compares LM, fE$_{50}$, fE$_{100}$, fE$_{200}$, fE$_{400}$, fE$_{600}$, fRE$_{50}$, fRE$_{100}$, fRE$_{200}$, fRE$_{400}$, and fRE$_{600}$, where subscripts represent $L$, by performing simulations under $n \in$ {5,000, 10,000, 20,000, 40,000, 80,000}.

Simulated data are generated with Eq. (25), whose eigen-pairs {**E**, **Λ**}, which are computationally intractable when $n$ is large, are replaced with approximated eigen-pairs {$\hat{\mathbf{E}}$ and $\hat{\mathbf{\Lambda}}$} that are extracted from $\mathbf{C}_L$, whose ($I$, $J$)-th element is given by exp(-$d(s_I, s_J)/r_{true}$), where $r_{true}$ denotes the true range parameter. We assume three cases that the true range parameter is half of, equal to, and twice the pre-specified value. In other words, $r_{true}$ is given by $r_{true} \in \{0.5r, r, 2.0r\}$, where $r$ is estimated a priori as explained in Section 2.1.

True values for the other parameters are the same as those in the previous section, except that $\sigma_\gamma = 1.0$. In other words, the models are estimated 200 times while varying $n$, $\sigma_{\gamma(x(k))} \in \{0, 0.6\}$, and $r_{true} \in \{0.5r, r, 2.0r\}$.

6.2. Result

Results for biases are not shown because they are almost zero across cases. Figure 5 plots the estimated RMSEs. The RMSEs of the LM are small when $\sigma^2_{\gamma(x)} = 0$. By contrast, the RMSEs increase when $\sigma^2_{\gamma(x)} = 0.6$. This result suggests that LM estimates are erroneous when explanatory variables are spatially dependent. By contrast, the RMSEs of the fast ESF and RE-ESF are nearly zero in all cases. This finding holds even if the number of eigenvectors, *L*, is small.

**[Figure 5 around here]**

Figure 6 plots the RMSPEs of the coefficient standard errors. RMSPEs for the ESF/RE-ESF are smaller than those for the LM. This outcome confirms that consideration of spatial dependence is needed even for large samples to evaluate coefficient standard errors accurately. An interesting finding is that the RMSPE of the LM decreases when $\mathbf{x}_1$ and $\mathbf{x}_2$ are spatially dependent (i.e., $\sigma^2_{\gamma(x)} = 0.6$). This might be because influences from residual spatial variations are partially absorbed by the spatial component in the regression term.

**[Figure 6 around here]**

With regard to ESF and RE-ESF, their RMSPEs are very close to each other,

although that for the RE-ESF is slightly better than that for the ESF when $\sigma^2_{\gamma(x)} = 0.6$.
Also conceivable is that the RMSPEs for ESF and RE-ESF increase when residuals are locally spatially dependent (i.e., $r_{true} = 0.5$). Such an outcome arises because the $L$ eigenvectors describe the $L$ most global spatial variations, which are explained by the MC. Nevertheless, their RMSPEs are small if $L$ is not too small, say $L \geq 200$. For example, the RMSPEs for ESF and RE-ESF are less than 0.070 across all cases, whereas those for LM are a maximum of 0.415. Although the range parameter $r$ is misspecified when $r_{true}$ = 0.5 or 2.0, increasing errors due to this misspecification are inconceivable. ESF and RE-ESF are verified to be robust against this misspecification. For ESF, such robustness might be because parameter estimation does not use eigenvalues in which $r$ determines the decay pattern. For RE-ESF, it is because the parameter $\alpha$ instead of $r$ estimates the scale (see Figure 1).

To examine if positive spatial dependence in residuals is successfully filtered, the z-value of the residual MC, $z[MC]$, is evaluated. Approximately, residual spatial dependence is statistically significant at the 5% level when $z[MC] > 1.96$. Table 5 summarizes $z[MC]$s for the LM, This table simply shows the existence of strong spatial dependence in the residuals. Figure 7 displays $z[MC]$s for the ESF and RE-ESF models, The figure shows that use of these models drastically reduces residual spatial dependence.

Increasing *L* makes the reduction more pronounced. Roughly speaking, $z[MC]$ is sufficiently small when *L* equals 200 or more across sample sizes. Appendix A analytically justifies this simulation result.

[Table 5 around here]

[Figure 7 around here]

Finally, Figure 8 compares computational times. When *L* is large, computational time rapidly increases as *n* grows. In contrast, the increase is very small when $L \leq 200$. Based on results in this section, $fE_{200}$ or $fRE_{200}$ would be sensitive choices to estimate regression coefficients accurately and computationally efficiently.

[Figure 8 around here]

In addition to the experiments discussed above, we also performed (i) simulations when the exponential kernel in $\mathbf{C}_L$ is replaced with other kernels, and (ii) simulations when the true value for the scale parameter *α* is changed. The former confirms that our approach accurately estimates regression coefficients even if the kernel function is changed. The latter demonstrates that the estimates are more accurate when residuals are globally spatially dependent. The result is consistent with the finding that errors decrease when *r* is large. See Appendices B and C for further detail.

## 7. Concluding remarks

This study develops the fast ESF and RE-ESF approaches for large spatial data, and reveals that both of these approaches accurately estimate regression coefficients with computational efficiency. These findings are meaningful because computational complexity is one of the biggest drawbacks of ESF (Dormann et al., 2007).

The fast ESF, whose model is identical to the LM, does not require stepwise eigenvector selection. The fast ESF is easily extended to non-Gaussian models, non-linear models, and many others (see Griffith, 2002, 2004b), by simply introducing the approximated *L* eigenvectors for explanatory variables, although $L = 200$ may not be the best even in these models.

The fast RE-ESF model also can be extended to models whose likelihood function is identical to Eq. (20). Such models include the RE-ESF-based spatially varying coefficients model (Murakami et al., 2017), and the RE-ESF with another random effects term, such as group effects. Bates (2007, 2010) extends a linear mixed effects model, which is identical to our model, to non-Gaussian/non-linear (non-spatial) mixed effects models. Extension of our approach to non-Gaussian and/or non-linear modeling while maintaining the computational efficiency is an important next step.

Consideration of negative spatial dependence is another remaining important issue (Griffith, 2006; Griffith and Arbia, 2010). Unfortunately, our Nyström extension-based approach is a smoothing approach that is not suitable for capturing negative spatial dependence. Integration of our proposed model with another model describing negative spatial dependence might be a possible approach to accommodate negative spatial dependence.

Furthermore, to reveal advantages and disadvantages of our approach, it must be compared with other fast approximations, which are listed in Section 1. Comparisons with the generalized method of moment (GMM)-based spatial regression, which is a well-known fast approach in spatial econometrics (Kelejian and Prucha, 1998; 1999; 2010), also would be beneficial for clarifying in which case our approach should be used.

We focus on "large sample size," which is an aspect of recent spatial data. Yet, recent data, which are typically collected through sensors, also contain observation error, location error, and sampling bias due to concealment processing (see. Arbia, Espa, and Giuliani, 2016). Another important future research effort is to extend the fast ESF and RE-ESF approaches to large and noisy spatial data.

The fast ESF and RE-ESF are implemented in an R package "spmoran" (https://cran.r-project.org/web/packages/spmoran/index.html). See Murakami (2017) for illustration.

Table 1: A comparison of the LM, ESF, and RE-ESF models

| Model | Eigen-decomposition | Candidate eigenvectors | Selection |
|---|---|---|---|
| LM | N.A. | N.A. | N.A. |
| E$_{step}$ | Exact | $\lambda_l > 0$ | Forward stepwise |
| E$_{lasso}$ | | | LASSO |
| RE | | | |
| fE$_{100}$ | Nyström extension | $\hat{\lambda}_l > 0 \mid l \in \{1, \cdots 100\}$ | All |
| fE$_{200}$ | | $\hat{\lambda}_l > 0 \mid l \in \{1, \cdots 200\}$ | |
| fE$_{100*}$ | | $\hat{\lambda}_l > 0 \mid l \in \{1, \cdots 100\}$ | $Cor(\mathbf{y}, \hat{\mathbf{e}}_l) > 0.01$[1] |
| fE$_{200*}$ | | $\hat{\lambda}_l > 0 \mid l \in \{1, \cdots 200\}$ | |
| fRE$_{50}$ | | $\hat{\lambda}_l > 0 \mid l \in \{1, \cdots 50\}$ | |
| fRE$_{100}$ | | $\hat{\lambda}_l > 0 \mid l \in \{1, \cdots 100\}$ | All |
| fRE$_{200}$ | | $\hat{\lambda}_l > 0 \mid l \in \{1, \cdots 200\}$ | |

[1] 0.2, 0.1, and 0.01 are tested for the threshold. The result shows that 0.01 is the best in terms of the estimation error of regression coefficients.

Table 2: The Bias of $\beta_1$. Darker cell denotes larger bias in absolute value. Greater $\sigma_{\gamma(x(k))}$ and $\sigma_\gamma$ mean stronger spatial dependent variations in explanatory variables and residuals, respectively.

|  | Bias | | | | | | Bias/Bias(LM) | | | | | |
|---|---|---|---|---|---|---|---|---|---|---|---|---|
| $\sigma_{\gamma(x(k))}$ | 0.0 | | | 0.6 | | | 0.0 | | | 0.6 | | |
| $\sigma_\gamma$ | 0.5 | 1.0 | 2.0 | 0.5 | 1.0 | 2.0 | 0.5 | 1.0 | 2.0 | 0.5 | 1.0 | 2.0 |
| LM | -0.001 | -0.001 | 0.001 | 0.010 | -0.012 | -0.008 | 1.00 | 1.00 | 1.00 | 1.00 | 1.00 | 1.00 |
| $E_{step}$ | -0.004 | -0.003 | -0.002 | -0.003 | -0.002 | 0.001 | 5.81 | 6.81 | 4.47 | 0.27 | 0.12 | 0.18 |
| $E_{lasso}$ | -0.004 | -0.003 | 0.002 | 0.007 | -0.002 | -0.001 | 4.91 | 6.89 | 4.27 | 0.70 | 0.15 | 0.07 |
| $fE_{100}$ | -0.002 | 0.000 | 0.001 | 0.003 | 0.002 | 0.002 | 2.15 | 0.39 | 1.52 | 0.34 | 0.17 | 0.27 |
| $fE_{200}$ | -0.002 | 0.001 | 0.001 | 0.003 | 0.001 | -0.002 | 2.88 | 1.54 | 1.70 | 0.33 | 0.08 | 0.24 |
| $fE_{100*}$ | -0.003 | 0.002 | -0.002 | -0.019 | -0.034 | -0.031 | 4.69 | 3.37 | 3.60 | 1.92 | 2.78 | 3.69 |
| $fE_{200*}$ | -0.007 | -0.002 | -0.005 | -0.024 | -0.046 | -0.056 | 9.54 | 3.19 | 9.64 | 2.43 | 3.72 | 6.69 |
| RE | -0.002 | 0.001 | 0.001 | 0.004 | 0.001 | -0.002 | 2.39 | 1.00 | 1.47 | 0.35 | 0.08 | 0.21 |
| $fRE_{50}$ | -0.001 | -0.001 | 0.002 | 0.003 | 0.001 | 0.011 | 2.02 | 1.10 | 3.35 | 0.27 | 0.05 | 1.36 |
| $fRE_{100}$ | -0.002 | 0.000 | 0.000 | 0.003 | 0.003 | 0.001 | 2.10 | 0.75 | 0.79 | 0.25 | 0.24 | 0.10 |
| $fRE_{200}$ | -0.002 | 0.000 | 0.001 | 0.004 | 0.001 | -0.002 | 2.38 | 0.50 | 1.02 | 0.35 | 0.09 | 0.27 |

Table 3: The RMSE of $\beta_1$. Darker cell denotes larger RMSE. Note: See Table 2

| | RMSE | | | | | | RMSE/RMSE(LM) | | | | | |
|---|---|---|---|---|---|---|---|---|---|---|---|---|
| $\sigma_{\gamma(x(k))}$ | 0.0 | | | 0.6 | | | 0.0 | | | 0.6 | | |
| $\sigma_\gamma$ | 0.5 | 1.0 | 2.0 | 0.5 | 1.0 | 2.0 | 0.5 | 1.0 | 2.0 | 0.5 | 1.0 | 2.0 |
| LM | 0.014 | 0.023 | 0.030 | 0.087 | 0.161 | 0.368 | 1.00 | 1.00 | 1.00 | 1.00 | 1.00 | 1.00 |
| E$_{step}$ | 0.015 | 0.015 | 0.017 | 0.027 | 0.027 | 0.026 | 1.06 | 0.67 | 0.57 | 0.30 | 0.17 | 0.07 |
| E$_{lasso}$ | 0.013 | 0.013 | 0.018 | 0.025 | 0.024 | 0.026 | 0.94 | 0.59 | 0.59 | 0.28 | 0.15 | 0.07 |
| fE$_{100}$ | 0.013 | 0.015 | 0.017 | 0.024 | 0.027 | 0.037 | 0.96 | 0.66 | 0.55 | 0.28 | 0.17 | 0.10 |
| fE$_{200}$ | 0.013 | 0.015 | 0.015 | 0.025 | 0.026 | 0.027 | 0.97 | 0.64 | 0.52 | 0.29 | 0.16 | 0.07 |
| fE$_{100*}$ | 0.015 | 0.015 | 0.015 | 0.038 | 0.060 | 0.054 | 1.06 | 0.65 | 0.51 | 0.44 | 0.37 | 0.15 |
| fE$_{200*}$ | 0.015 | 0.015 | 0.014 | 0.040 | 0.059 | 0.080 | 1.12 | 0.68 | 0.48 | 0.45 | 0.37 | 0.22 |
| RE | 0.013 | 0.015 | 0.016 | 0.023 | 0.023 | 0.026 | 0.97 | 0.65 | 0.52 | 0.26 | 0.14 | 0.07 |
| fRE$_{50}$ | 0.014 | 0.016 | 0.018 | 0.023 | 0.041 | 0.067 | 0.99 | 0.70 | 0.60 | 0.27 | 0.25 | 0.18 |
| fRE$_{100}$ | 0.013 | 0.014 | 0.016 | 0.023 | 0.027 | 0.038 | 0.97 | 0.64 | 0.54 | 0.26 | 0.17 | 0.10 |
| fRE$_{200}$ | 0.013 | 0.015 | 0.016 | 0.023 | 0.023 | 0.027 | 0.97 | 0.64 | 0.52 | 0.26 | 0.14 | 0.07 |

Table 4: The RMSPE of $se(\beta_1)$. Darker cell denotes larger RMSPE. Note: See Table 2

|  | RMSPE | | | | | | RMSPE/RMSPE(LM) | | | | | |
|---|---|---|---|---|---|---|---|---|---|---|---|---|
| $\sigma_{\gamma(x(k))}$ | 0.0 | | | 0.6 | | | 0.0 | | | 0.6 | | |
| $\sigma_\gamma$ | 0.5 | 1.0 | 2.0 | 0.5 | 1.0 | 2.0 | 0.5 | 1.0 | 2.0 | 0.5 | 1.0 | 2.0 |
| LM | 0.110 | 0.395 | 1.196 | 0.302 | 0.188 | 0.224 | 1.00 | 1.00 | 1.00 | 1.00 | 1.00 | 1.00 |
| E$_{step}$ | 0.006 | 0.006 | 0.006 | 0.072 | 0.063 | 0.048 | 0.05 | 0.02 | 0.00 | 0.24 | 0.34 | 0.21 |
| E$_{lasso}$ | 0.004 | 0.003 | 0.003 | 0.062 | 0.036 | 0.020 | 0.04 | 0.01 | 0.00 | 0.21 | 0.19 | 0.09 |
| fE$_{100}$ | 0.009 | 0.026 | 0.105 | 0.076 | 0.026 | 0.054 | 0.08 | 0.07 | 0.09 | 0.25 | 0.14 | 0.24 |
| fE$_{200}$ | 0.013 | 0.013 | 0.032 | 0.129 | 0.053 | 0.031 | 0.12 | 0.03 | 0.03 | 0.43 | 0.28 | 0.14 |
| fE$_{100*}$ | 0.008 | 0.026 | 0.120 | 0.054 | 0.025 | 0.057 | 0.08 | 0.07 | 0.10 | 0.18 | 0.13 | 0.26 |
| fE$_{200*}$ | 0.008 | 0.012 | 0.043 | 0.095 | 0.029 | 0.022 | 0.07 | 0.03 | 0.04 | 0.31 | 0.15 | 0.10 |
| RE | 0.009 | 0.014 | 0.019 | 0.026 | 0.011 | 0.014 | 0.08 | 0.04 | 0.02 | 0.09 | 0.06 | 0.06 |
| fRE$_{50}$ | 0.011 | 0.061 | 0.238 | 0.045 | 0.052 | 0.096 | 0.10 | 0.16 | 0.20 | 0.15 | 0.28 | 0.43 |
| fRE$_{100}$ | 0.004 | 0.018 | 0.095 | 0.018 | 0.024 | 0.042 | 0.04 | 0.05 | 0.08 | 0.06 | 0.13 | 0.19 |
| fRE$_{200}$ | 0.008 | 0.009 | 0.015 | 0.022 | 0.013 | 0.014 | 0.08 | 0.02 | 0.01 | 0.07 | 0.07 | 0.06 |

Table 5: z-value of residual MC (LM)

| | Sample size | 5,000 | 10,000 | 20,000 | 40,000 | 80,000 |
|---|---|---|---|---|---|---|
| $\sigma_{\gamma(x)}$ | 0.0 | 462 | 783 | 1,308 | 2,410 | 5,736 |
| | 0.6 | 358 | 556 | 1,192 | 2,584 | 6,433 |

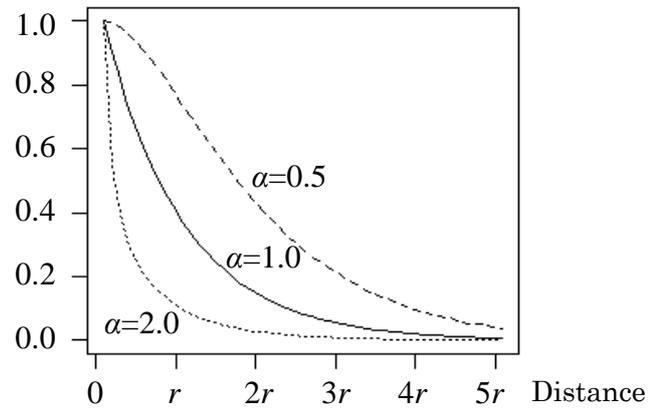

Figure 1: The scale parameter $\alpha$ and the distance decay of the elements in $\mathbf{C}^\alpha = \mathbf{E}\boldsymbol{\Lambda}(\alpha)\mathbf{E}'$, where $r$ is the range parameter in $\mathbf{C}$.

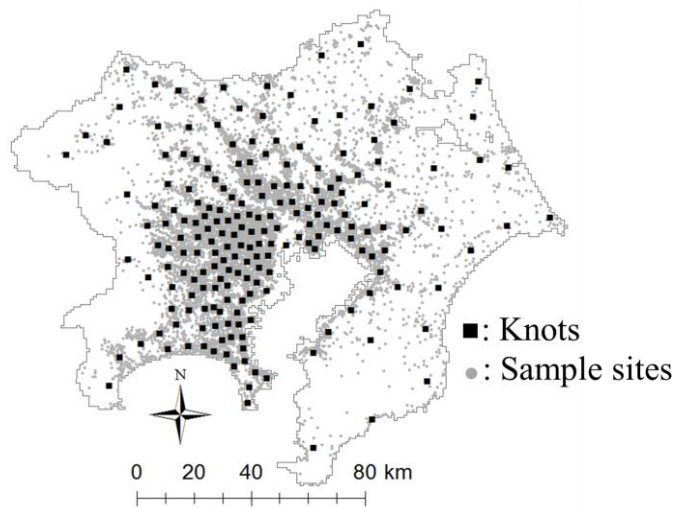

Figure 2: 200 knots extracted from the 9,215 official assessment sites on residential land prices in 2010 in the Tokyo metropolitan area.

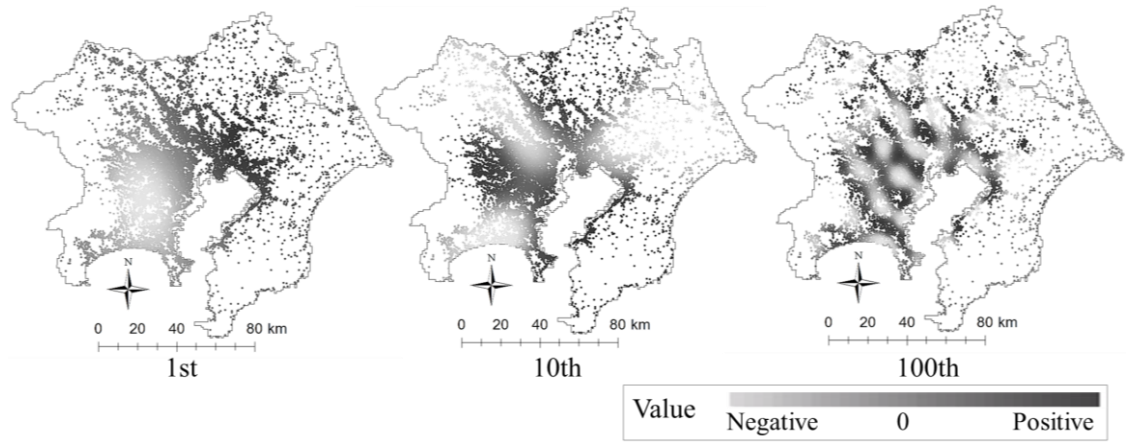

Figure 3: The approximated 1st, 10th, and 100th eigenvectors

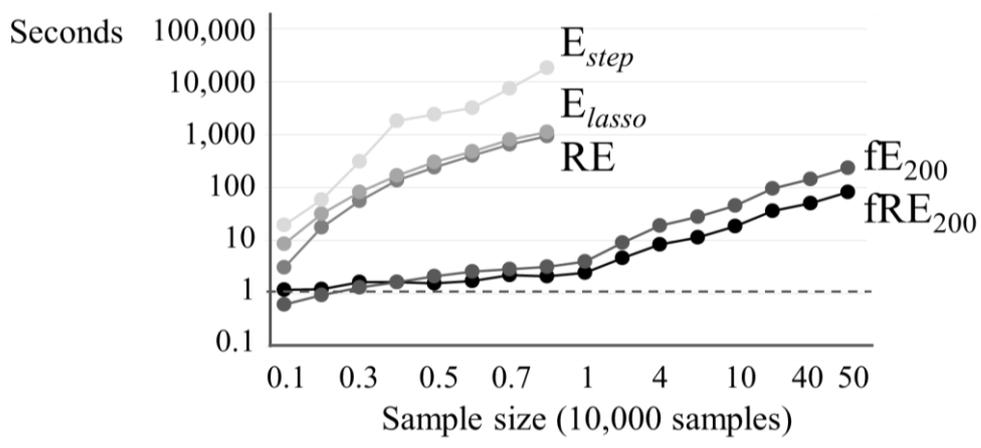

Figure 4: Computational time (selected ESF and RE-ESF specifications)

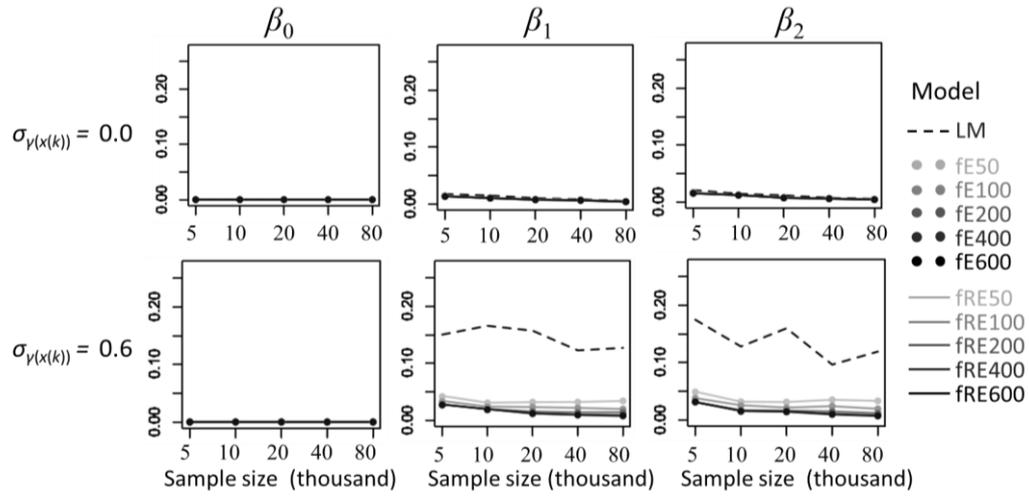

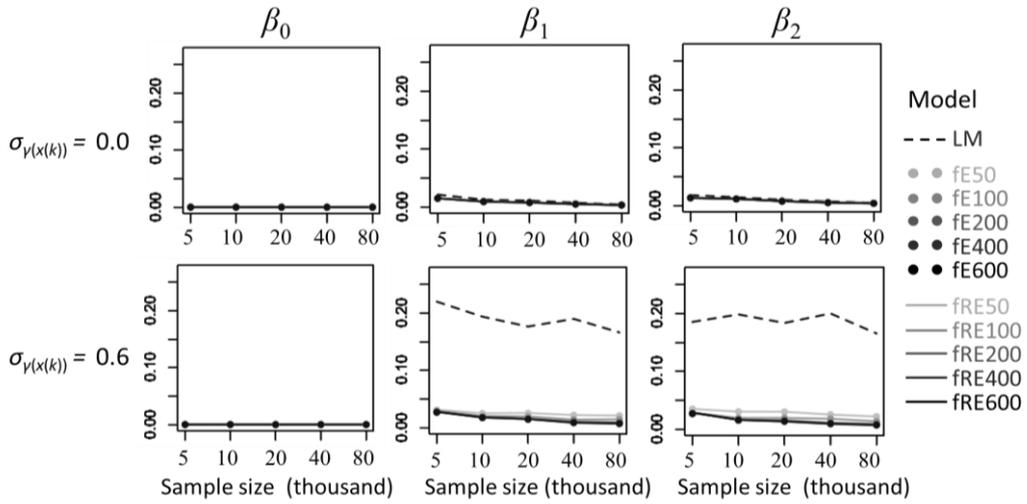

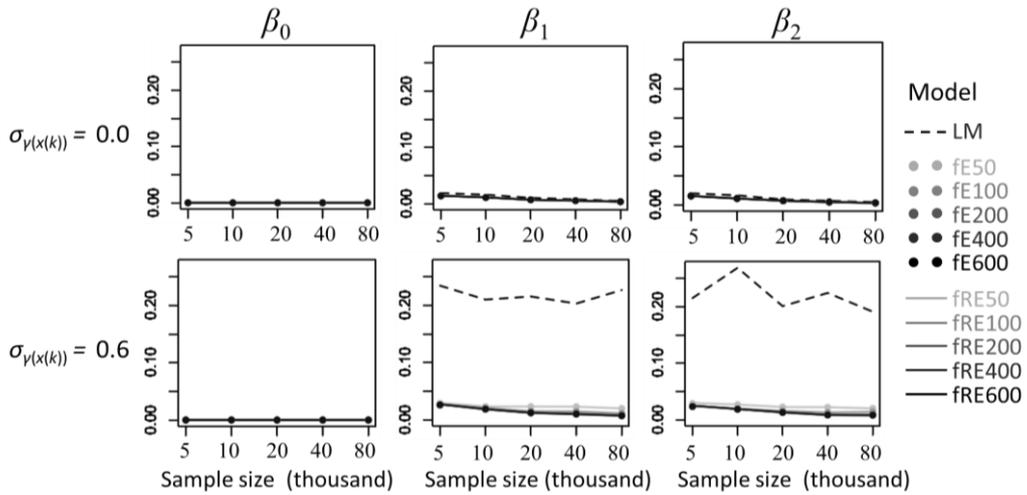

Figure 5: RMSE of regression coefficients

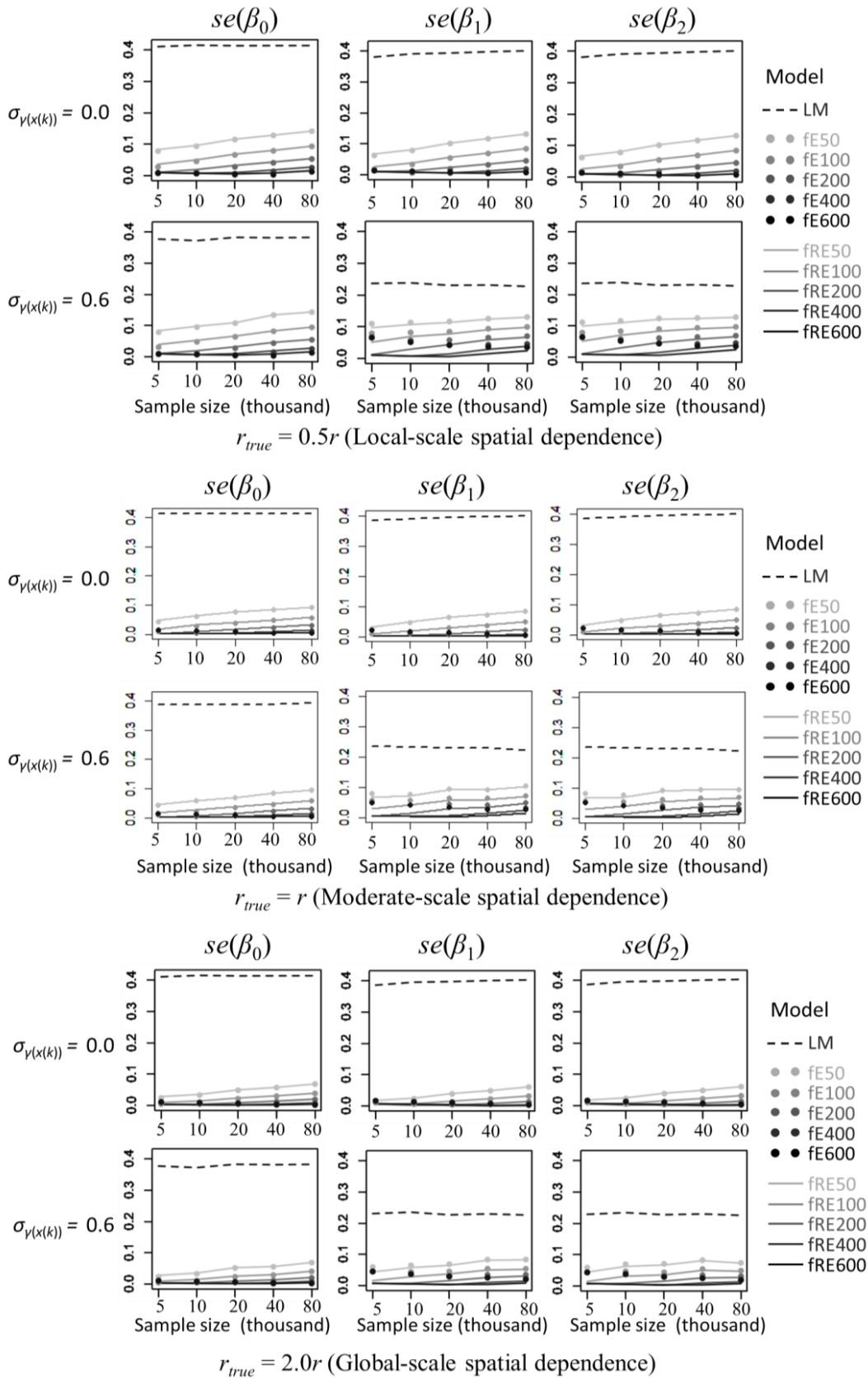

Figure 6: RMSPE of coefficients standard errors

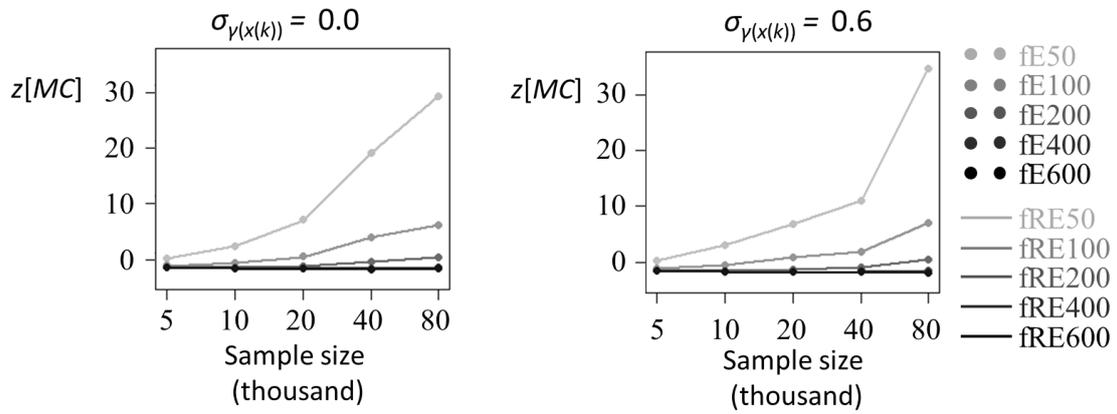

Figure 7: z-value of the residual MC (fast ESF and RE-ESF models)

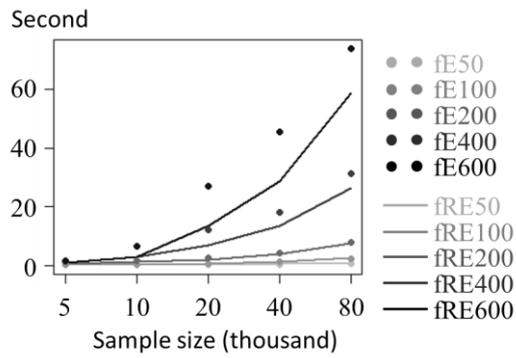

Figure 8: Computational time

**Appendix A. Analytic result on Moran's eigenvectors and spatial dependence**

ESF typically eliminates positive spatial dependent (P-SD) variations in residuals using the first *L* Moran's eigenvectors. This appendix analytically evaluates the ratio of P-SD variations being explained by the *L* eigenvectors while varying the *L* value. The cumulative contribution ratio, $p_l = \sum_{l=1}^{L} \lambda_l \Big/ \sum_{l=1}^{L^P} \lambda_l$, where $L^P$ is the number of positive eigenvalues, is used for the evaluation. $p_l = 1$ suggests that P-SD variations are fully captured by the *L* eigenvectors. We seek reasonably small *L* that achieves large $p_l$ (i.e., near 1). Throughout this appendix, we assume a 2-dimensional space that is consists of $N \times N$ sample sites regularly arranged with intervals of $1/2\pi$. It is also assumed that the matrix elements of **C** ($N^2 \times N^2$) are given by $\exp(-d_{i,j}/r)$ following our assumption.

The eigenvalues of **C** + **I** have the following analytic solutions:

$$\lambda_l^{(\mathbf{C}+\mathbf{I})} = n \frac{\tau_l}{\sum_{l=1}^{n} \tau_l} \tag{A1}$$

$$\tau_l = \left(\frac{1}{r^2} + 4\pi^2 l^2\right)^{-1.5} \tag{A2}$$

where $n = N^2$. (A1) represents a constraint that the sum of the eigenvalues equals $n$ (= the trace of **C** + **I**). Eq.(A2) is explained in Rasmussen and Williams (2006). Following Section 2, *r* is given by the maximum length of the minimum spanning tree connecting sample sites that always equals $1/2\pi$ on the regular grid.

The eigenvalues of **C** is readily given from Eq.(A1) as follows:

$$\lambda_l^{(\mathbf{C})} = n \frac{\tau_l}{\sum_{l=1}^{n} \tau_l} - 1 \tag{A3}$$

Honeine (2014) derived the following relationship between $\lambda_l^{(\mathbf{C})}$ and the eigenvector of **MCM**, $\lambda_l$:

$$\lambda_{l+1}^{(\mathbf{C})} \leq \lambda_l \leq \lambda_l^{(\mathbf{C})} \tag{A4}$$

Eq.(A4) indicates that $L^P \leq L^{(\mathbf{C})P}$, where $L^{(\mathbf{C})P}$ is the number of positive eigenvalues of **C**. He also derived the following inequality:

$$\sum_{l}^{L} \left[ \lambda_l^{(\mathbf{C})} + \left( \frac{\mathbf{1'C1}}{n} - \frac{2}{n} \lambda_l^{(\mathbf{C})} \right) (\mathbf{e}_l^{(\mathbf{C})'} \mathbf{1})^2 \right] \leq \sum_{l}^{L} \lambda_l \tag{A5}$$

which is readily expanded using $L^P \leq L^{(\mathbf{C})P}$ as

$$\frac{\sum_{l}^{L} \left[ \lambda_l^{(\mathbf{C})} + \left( \frac{\mathbf{1'C1}}{n} - \frac{2}{n} \lambda_l^{(\mathbf{C})} \right) (\mathbf{e}_l^{(\mathbf{C})'} \mathbf{1})^2 \right]}{\sum_{l=1}^{L^{(\mathbf{C})P}} \lambda_l} \leq \frac{\sum_{l=1}^{L} \lambda_l}{\sum_{l=1}^{L^{(\mathbf{C})P}} \lambda_l} \leq \frac{\sum_{l=1}^{L} \lambda_l}{\sum_{l=1}^{L^P} \lambda_l} = p_l \tag{A6}$$

where $\mathbf{e}_l^{(\mathbf{C})}$ is the $l$-th eigenvector of **C**. The eigenvectors on the regular grids are analytically given by sinusoidal. We use this property to evaluate $\mathbf{e}_l'\mathbf{1}$ for large $N$. Based on Eq.(A6), the left hand side of the equation is the lower bound of $p_l$.

Figure A plots the lower bound in cases with $L \in \{50, 100, 200, 400, 600\}$. This plots demonstrates that most of P-SD variations are explained even if $L$ is far more smaller than the sample size. For instance, when the sample size is 250,000 (= $500^2$), 88 % of P-SD variations are explained with only 200 eigenvectors. The result highlights the parsimony of ESF and RE-ESF approaches. These results are consistent with discussions in Section 6 showing that residual spatial dependence is effectively eliminated when $L$ is reasonably small, say 200.

Although it is possible to derive some criteria to determine $L$ based on the analytic results, analytic solution changes depending on (i) the kernel function and (ii) the value of the range parameter $r$. The former restricts the selection of the kernel function. The latter imposes us estimating $r$, which is slow for large $n$, a priori. Hence, we prefer $L$ = 200, which is readily available, rather than analytic criteria.

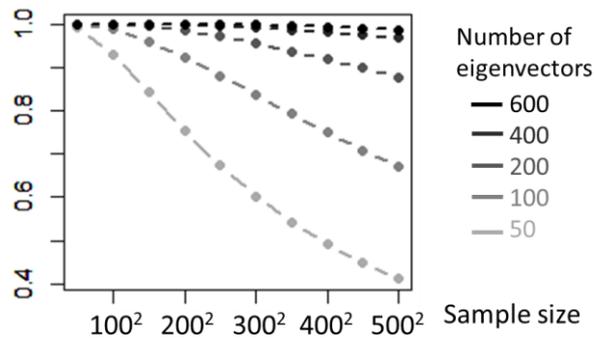

Figure A: Ratio of spatially dependent variations being explained by the eigenvectors

**Appendix B. Simulation experiments with different type of kernels**

This section performs a simulation experiment to examine the influence of the kernel selection on coefficient estimates. The experimental design is the same with the experiment in Section 6 except that $r_{true} = r$ is assumed for simplicity, and that the exponential kernel is replaced with the spherical kernel and the Gaussian kernel, which are widely used in geostatistics..

Bias and RMSE of their coefficients are almost zero irrespective of the kernel function. RMSPE of coefficients standard errors when $\sigma_{\gamma(x(k))} = 0.0$ are plotted in Figure B. The RMSPEs are small when the Gaussian kernels are used. This is because the Gaussian kernel describes a smoother spatial process than the other two (see, Cressie, 1993) whereas accuracy of low rank models including our models are usually increase when the process is smooth. Because the spherical kernel describes a rather spatial process, the RMSPEs slightly increases relatively the exponential kernel. Still, the RMSPEs are quite small relative to LM especially when $L \geq 200$. It is confirmed that fE200 and fRE200 accurately estimates parameters even if their kernel is changed.

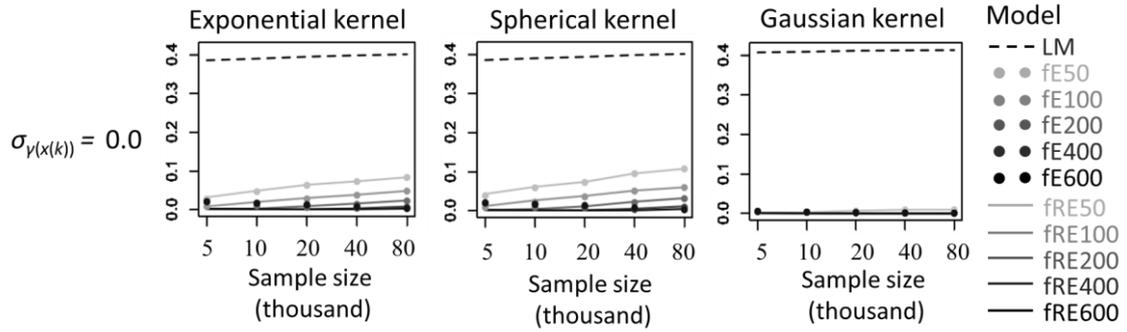

Figure B: Kernels and RMSPE of the standard errors of $\beta_1$.

**Appendix C. Simulation experiments on the parameter $\alpha$**

This section performs a simulation experiment to examine the influence of the value of $\alpha$ on coefficient estimates. The simulation design is the same with Section 6 except that $\alpha \in \{0.5, 1.0, 2.0\}$ is and $r_{true}$ is fixed by $r$.

The RMSEs of the estimated coefficients and the RMSPEs of their standard errors are summarized in Figures C1 and C2, respectively. These Figures show that the RMSPEs increase when $\alpha = 0.5$. The result is consistent with the finding that the RMSPEs inflates when the range parameter $r$ is small (see Section 6). Extension of the fast ESF and RE-ESF to capture local-scale spatial dependence is an important next step.

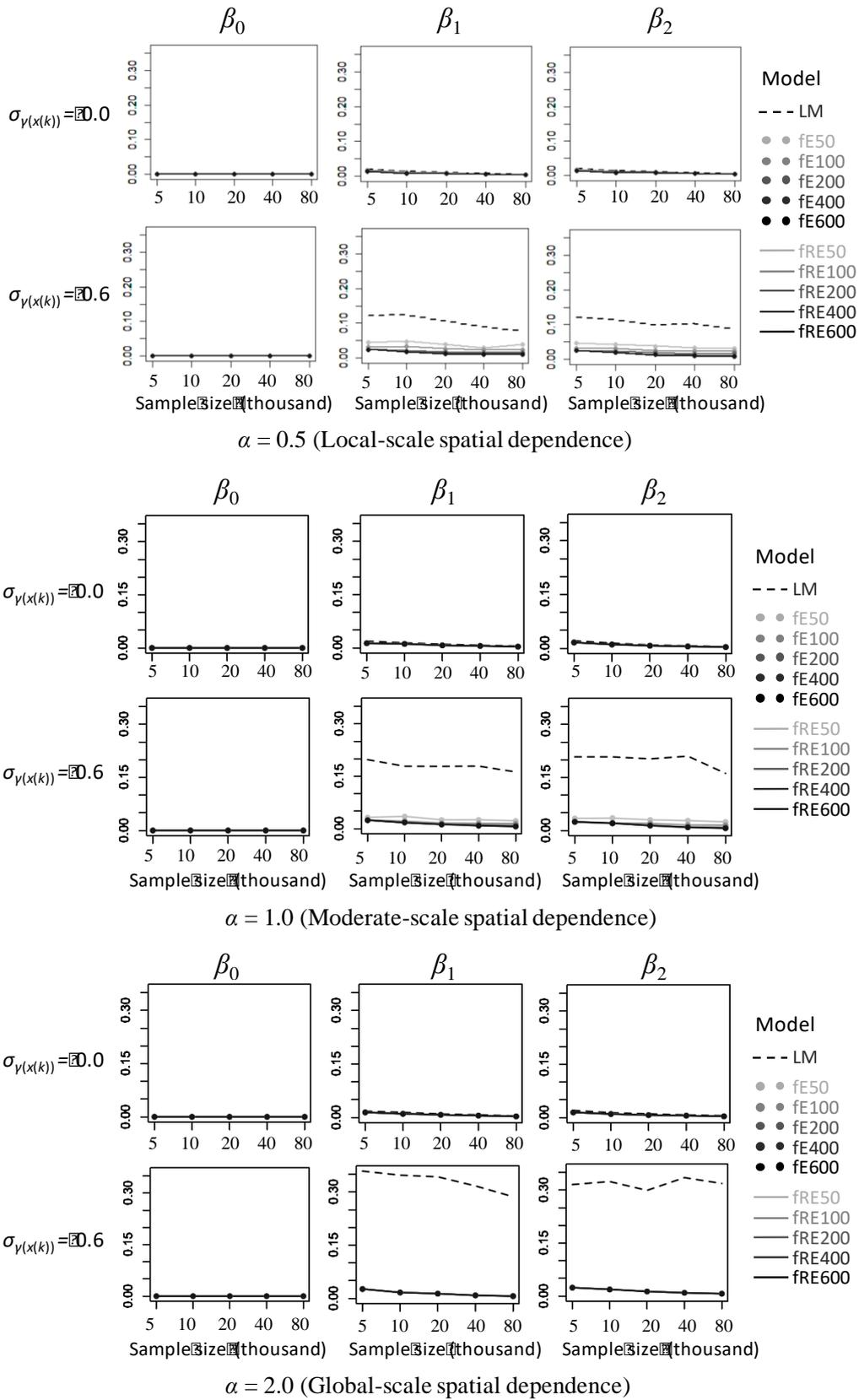

Figure C1: RMSE of regression coefficients when $\alpha$ = 0.5, 1.0, and 2.0

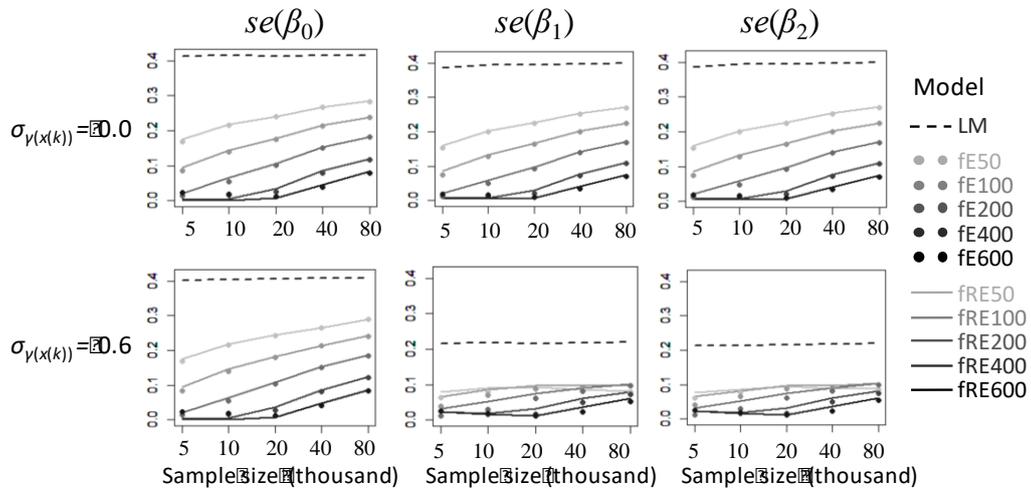

$\alpha = 0.5$ (Local-scale spatial dependence)

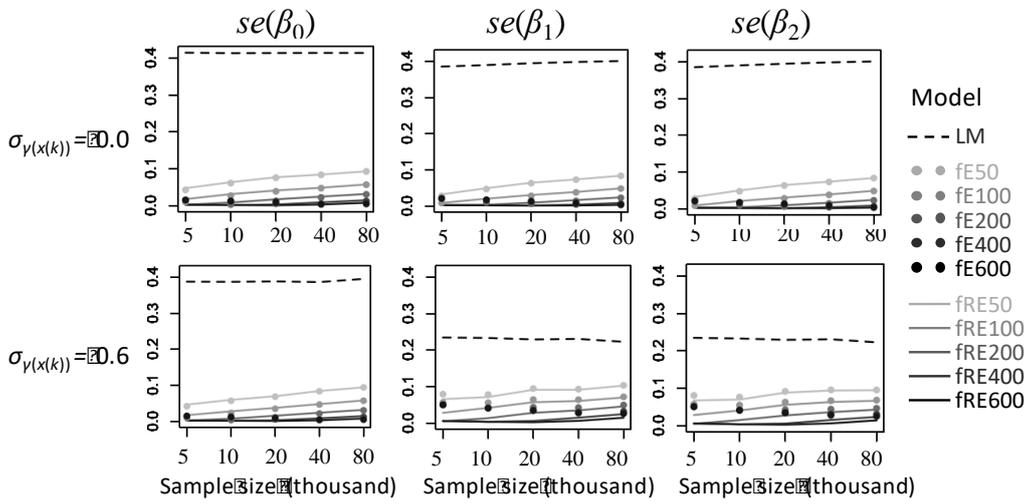

$\alpha = 1.0$ (Moderate-scale spatial dependence)

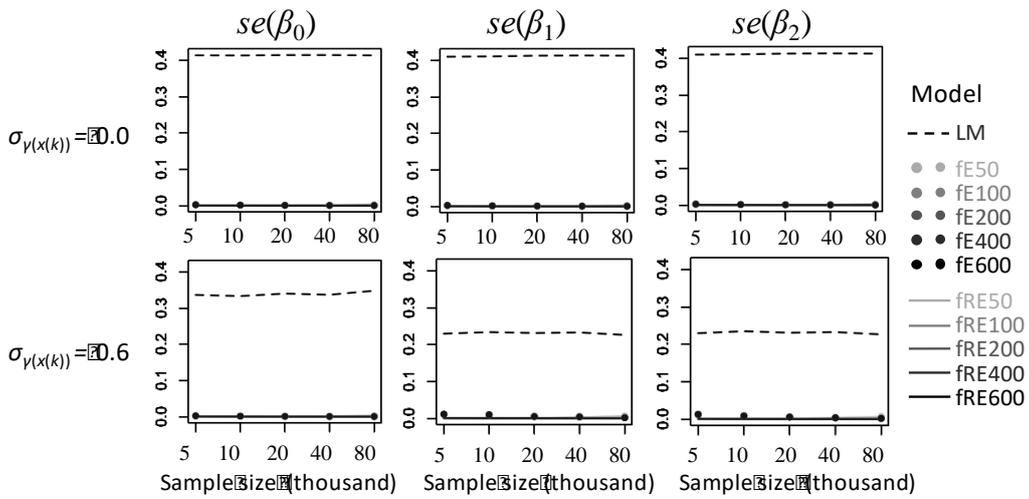

$\alpha = 2.0$ (Global-scale spatial dependence)

Figure C2: RMSPE of coefficients standard errors when $\alpha$= 0.5, 1.0, and 2.0